**Do Electric Vehicles Induce More Motion Sickness Than Fuel Vehicles? A Survey Study in China**

**Weiyin Xie**
Thrust of Robotics and Autonomous Systems, Systems Hub
The Hong Kong University of Science and Technology (Guangzhou), Guangzhou, China
Email: wxie593@connect.hkust-gz.edu.cn

**Chunxi Huang**
Robotics and Autonomous Systems, Interdisciplinary Programs Office
The Hong Kong University of Science and Technology, Hong Kong SAR, China, 999077
Email: tracy.huang@connect.ust.hk

**Jiyao Wang**
Thrust of Intelligent Transportation, Systems Hub
The Hong Kong University of Science and Technology (Guangzhou), Guangzhou, China
Email: jwang297@connect.hkust-gz.edu.cn

**Dengbo He[*]**
Thrust of Intelligent Transportation, Systems Hub
The Hong Kong University of Science and Technology (Guangzhou), Guangzhou, China
Department of Civil and Environmental Engineering
The Hong Kong University of Science and Technology, Hong Kong SAR, China
Email: dengbohe@hkust-gz.edu.cn

[*] Corresponding author.

Word Count: 5,458 words + 3 table (250 words per table) = 6,208 words

*Submitted Aug 1st, 2023*

*Weiyin Xie, Chunxi Huang, Jiyao Wang, and Dengbo He*

**ABSTRACT**
Electric vehicles (EVs) are a promising alternative to fuel vehicles (FVs), given some unique characteristics of EVs, for example, the low air pollution and maintenance cost. However, the increasing prevalence of EVs is accompanied by widespread complaints regarding the high likelihood of motion sickness (MS) induction, especially when compared to FVs, which has become of the major obstacles to the acceptance and popularity of EVs. Despite the prevalence of such complaints online and among EV users, the association between vehicle type (i.e., EV versus FV) and MS prevalence and severity has not been quantified. Thus, this study aims to investigate the existence of EV-induced MS and explore the potential factors leading to it. A survey study was conducted to collect passengers' MS experience in EVs and FVs in the past one year. In total, 639 valid responses were collected from mainland China. The results show that FVs were associated with a higher frequency of MS, while EVs were found to induce more severe MS symptoms. Further, we found that passengers' MS severity was associated with individual differences (i.e., Age, gender, sleep habits, susceptibility to motion-induced MS), in-vehicle activities (i.e., chatting with others and watching in-vehicle displays), and road conditions (i.e., congestion and slope), while the MS frequency was associated with the vehicle ownership and riding frequency. The results from this study can guide the directions of future empirical studies that aim to quantify the inducers of MS in EVs and FVs, as well as the optimization of EVs to reduce MS.

*Weiyin Xie, Chunxi Huang, Jiyao Wang, and Dengbo He*

**INTRODUCTION**
The electric vehicle (EV) driven by the electric motor is one of the most promising alternatives to fuel vehicle (FV) driven by internal combustion engines. By 2035, the share of EVs in the Chinese automobile market is expected to reach 50% (*1*). EVs offer several advantages over traditional FVs, such as faster dynamic responses (*2*) of the power system and lower carbon emissions (*2*). However, users or potential users still have concerns over EVs, and motion sickness (MS) is one of them.

MS is a physical discomfort that is caused by the passive movement of the human body. It is characterized by symptoms such as nausea, dizziness, and vomiting (*3*). The MS has attracted attention from various fields, including aviation (*4*), rail transit (*5*), virtual reality (*6*), and vehicular control (*7*). Although research on MS in vehicles has been ongoing for decades, the deployment of new innovative technologies, such as autonomous driving (*8*), new human-machine interface designs (*9*), and electrification of vehicle power systems (*10*), may introduce new sources of MS in vehicles, which could hinder the widespread and acceptance of these new technologies. For example, a study on hybrid electric vehicle acceptance suggests that passenger comfort is a crucial factor influencing public acceptance (*11*), and MS induction can negatively impact the comfort of vehicles.

Understanding the causes of MS serves as the foundation to remove MS inducers. Several theories have been proposed to explain the causes of MS. For example, the theory of sensory conflict suggests that MS is induced by a mismatch between vestibular signals, visual signals, and somatosensory motion cues (*12*). When an occupant looks at an in-vehicle display, they may visually perceive themselves as in a static environment; while the vestibular signal and somatosensory motion perceived by their body suggest that they are in a moving environment. This conflict in perceived motion states can explain why vehicle-dynamics-related factors, such as the stop-and-go scenario caused by traffic jams (*13*), curvy roads (*13*), and slalom driving (*14*) can lead to more severe MS. Again, based on the theory of sensory conflict, previous research also suggests that the capability to anticipate future motion of vehicles can be associated with MS (*15*), given that knowing the future dynamics of a vehicle can reduce the perceived sensory conflict between the visual and vestibular channels. This could explain the higher incidence of MS in passengers than in drivers (*16*). At the same time, MS can also be induced by vibration alone. For example, previous research found that low frequency of vertical oscillation, low frequency of fore-and-aft oscillation, low frequency of lateral oscillation, and high lateral acceleration all may lead to MS in vehicles (*7*).

The characteristics of the EV are closely associated with the above-mentioned inducers of MS. On one hand, because of the fast dynamic response of the electrified drive system and the introduction of the energy regenerative system, EVs are usually associated with higher acceleration and unexpected deceleration. At the same time, the auditory cues from the internal combustion engine could be used to anticipate the acceleration of FVs, but they no longer exist in EVs as the electric motor can respond almost immediately to pedal movement. Thus, the EVs may lead to more severe MS compared to FVs, given the above-mentioned differences between the FVs and EVs. However, although there are numerous online complaints about EVs causing more MS than traditional FVs (*17*), no research has been conducted to verify whether this is the case or it is just another example of survivor bias (*18*), i.e., the majority of users who never experienced MS may tend to keep quiet. Without verifying the motion sickness severity in EVs, further endeavours to reduce motion sickness in EVs can be questionable.

To address this research gap, a survey study was conducted in mainland China, which is the largest EV market in the world (*19*). Given that MS is more prevalent among passengers (*16*), the survey targeted the population who took rides in both EVs and/or FVs as passengers in the past one year. Several scenario-related factors and demographic-related factors that may be associated with MS have also been investigated and the self-reported prevalence and the highest level of severity were used to evaluate the MS among passengers.

*Weiyin Xie, Chunxi Huang, Jiyao Wang, and Dengbo He*

**METHODS**
**Questionnaire design**
The questionnaire designed for data collection includes three parts: (1) Individual information; (2) MS susceptibility; (3) Experience of MS in EVs and FVs, which assessed respondents' frequency of experiencing MS in the past one year and the severity and scenario-related factors associated with their most recent experience of MS within the past one year. The questionnaire design is illustrated in **Table 1** and **Table 2**.

*Individual information*
As shown in **Table 1**, the respondents' individual information was collected to account for individual differences in experiencing MS. The basic demographic information was collected, including age, gender, workout routine, sleep pattern, smoking, and drinking habits. In addition, respondents' susceptibility to motion-induced MS was assessed using MSSQ-short (*20*). In the end, the ownership of EVs and FVs was also inquired. The citations after each variable provide related research that suggests the relationship between MS and these variables.

**Table 1: Individual-Related Questions and Extracted Variables.**

| Question | Variables | Type | Responses and Distribution |
|---|---|---|---|
| *Q1: What's your age?* | Age (*21*) | Continuous | - Mean: 28.5 (SD: 6.7, min: 17, max: 65) |
| *Q2: What's your gender?* | Gender (*22*) | Nominal | - Male (n=368, 55.1%)<br>- Female (n=295, 44.2%) |
| *Q3: How many hours per week have you exercised (including running, swimming, gym, etc.) in the past six months?* | Sport (*23*) | Ordinal | - Rare (n=64, 9.6%)<br>- Less than half hours (n=117, 17.6%)<br>- Half to one hour (n=159, 23.9%)<br>- One to two hours (n=122, 18.4%)<br>- Over two hours (n=201, 30.3%) |
| *Q4: How many hours do you usually sleep per day in the past six months?* | Sleep (*24*) | Ordinal | - Less than four hours (n=9, 1.3%)<br>- Four hours to six hours (n=68, 10.2%)<br>- Six hours to eight hours (n=509, 76.7%)<br>- Over eight hours (n=77, 11.6%) |
| *Q5: How many times per week on average have you consumed alcohol in the past six months?* | Alcohol (*25*) | Ordinal | - Rare (n=392, 59.1%)<br>- Less than one time (n=163, 24.5%)<br>- Two to four times (n=98, 14.7%)<br>- Over five times (n=10, 1.5%) |
| *Q6: How many cigarettes do you smoke per week?* | Smoke (*26*) | Ordinal | - Never (n=513, 77.3%)<br>- Less than five (n=34, 5.1%)<br>- Five to ten (n=41, 6.1%)<br>- Ten to twenty (n=26, 3.9%)<br>- Over twenty (n=49, 7.3%) |
| *Q7: Question sets from MSSQ questionnaire* | MSSQ (*20*) | Continuous | - Mean: 19.64 (SD: 12.6, min: 0.0, max: 54.0) |
| *Q8: Do you or your family own an EV* | EV ownership | Nominal | - Yes (215, 33.6%)<br>- No (424, 66.3%) |
| *Q9: Do you or your family own an FV* | FV ownership | Nominal | - Yes (308, 48.2%)<br>- No (331, 51.7%) |

**Table 2: MS-Related Questions and Extracted Variables.**

| Question | Variables | Type | Distribution | |
|---|---|---|---|---|
| | | | MS in EVs | MS in FVs |
| Q10: In the past one year, how often have you (as a passenger) taken a ride in an EV/FV? | Riding frequency (*15*) | Ordinal | - Almost every day (n=49, 7.6%)<br>- A few times a week (n=255, 39.9%)<br>- A few times a month (n=241, 37.7%)<br>- A few times a year (n=76, 11.8%)<br>- Never (n=18, 2.8%) | - Almost every day (n=62, 9.7%)<br>- A few times a week (n=237, 37.0%)<br>- A few times a month (n=250, 39.1%)<br>- A few times a year (n=75, 11.7%)<br>- Never (n=15, 2.3%) |
| Q11: In the past one year, in an EV/FV, how often have you experienced MS (as a passenger)? | MS frequency | Ordinal | - Always (n=41, 6.4%)<br>- Frequent (n=77, 12.0%)<br>- Occasional (n=282, 44.1%)<br>- Never (n=221, 34.5%)<br>- N/A (n=18, 2.8%) | - Always (n=50, 7.8%)<br>- Frequent (n=160, 25.0%)<br>- Occasional (n=237, 37.0%)<br>- Never (n=177, 27.6%)<br>- N/A (n=15, 2.3%) |
| Q12: In your most recent experience of motion sickness (as a passenger) in an EV/FV, how severe were your symptoms? | MS severity (*28*) | Ordinal | - Alright (n=41, 6.4%)<br>- Slightly unwell (n=251, 39.2%)<br>- Quite ill (n=65, 10.1%)<br>- Absolutely dreadful (n= 43, 6.7%)<br>- N/A (n=239, 37.4%) | - Alright (n=59, 9.2%)<br>- Slightly unwell (n=236, 36.9%)<br>- Quite ill (n=84, 13.1%)<br>- Absolutely dreadful (n= 66, 10.3%)<br>- N/A (n=194, 30.3%) |
| Q13: What descriptions match the behavior of the vehicle during your most recent occurrence of MS (as a passenger) in an EV/FV? | Acceleration and braking (*29*) | Nominal | - Yes (n=307, 48.0%)<br>- No (n=93, 14.5%)<br>- N/A (n=239, 37.4%) | - Yes (n=341, 53.3%)<br>- No (n=104, 16.2%)<br>- N/A (n=194, 30.3%) |
| | Turning (*13*) | Nominal | - Yes (n=236, 36.9%)<br>- No (n=164, 25.6%)<br>- N/A (n=239, 37.4%) | - Yes (n=282, 44.1%)<br>- No (n=163, 25.5%)<br>- N/A (n=194, 30.3%) |
| | Lane change (*14*) | Nominal | - Yes (n=185, 28.9%)<br>- No (n=215, 33.6%)<br>- N/A (n=239, 37.4%) | - Yes (n=231, 36.1%)<br>- No (n=214, 33.4%)<br>- N/A (n=194, 30.3%) |
| Q14: What descriptions match the road/traffic condition during your most recent occurrence of MS (as a passenger) in an EV/FV? | Congested road (*30*) | Nominal | - Yes (n=162, 25.3%)<br>- No (n=238, 37.2%)<br>- N/A (n=239, 37.4%) | - Yes (n=164, 25.6%)<br>- No (n=281, 43.9%)<br>- N/A (n=194, 30.3%) |
| | Smooth road (*30*) | Nominal | - Yes (n=32, 5.0%)<br>- No (n=368, 57.5%)<br>- N/A (n=239, 37.4%) | - Yes (n=35, 5.4%)<br>- No (n=410, 64.1%)<br>- N/A (n=194, 30.3%) |
| | Highway (*13*) | Nominal | - Yes (n=33, 5.1%)<br>- No (n=367, 57.4%)<br>- N/A (n=239, 37.4%) | - Yes (n=43, 6.7%)<br>- No (n=402, 62.9%)<br>- N/A (n=194, 30.3%) |
| | Bumpy Road (*31*) | Nominal | - Yes (n=187, 29.2%)<br>- No (n=213, 33.3%)<br>- N/A (n=239, 37.4%) | - Yes (n=207, 32.3%)<br>- No (n=238, 37.2%)<br>- N/A (n=194, 30.3%) |

| | Uphill Road (*32*) | Nominal | - Yes (n=25, 3.9%)<br>- No (n=375, 58.6%)<br>- N/A (n=239, 37.4%) | - Yes (n=34, 5.3%)<br>- No (n=411, 64.3%)<br>- N/A (n=194, 30.3%) |
|---|---|---|---|---|
| | Downhill Road (*32*) | Nominal | - Yes (n=31, 4.8%)<br>- No (n=369, 57.7%)<br>- N/A (n=239, 37.4%) | - Yes (n=38, 5.9%)<br>- No (n=407, 63.6%)<br>- N/A (n=194, 30.3%) |
| | Curvy road (*13*) | Nominal | - Yes (n=149, 23.3%)<br>- No (n=251, 39.2%)<br>- N/A (n=239, 37.4%) | - Yes (n=189, 29.5%)<br>- No (n=256, 40.0%)<br>- N/A (n=194, 30.3%) |
| *Q15: What activity were you engaged in inside the vehicle during your most recent experience of MS (as a passenger) in an EV/FV?* | Chatting (*13*) | Nominal | - Yes (n=74, 11.5%)<br>- No (n=326, 51.0%)<br>- N/A (n=239, 37.4%) | - Yes (n=87, 13.6%)<br>- No (n=358, 56.0%)<br>- N/A (n=194, 30.3%) |
| | Listening to audio (*13*) | Nominal | - Yes (n=85, 13.3%)<br>- No (n=315, 49.2%)<br>- N/A (n=239, 37.4%) | - Yes (n=83, 13.6%)<br>- No (n=362, 56.6%)<br>- N/A (n=194, 30.3%) |
| | Closing eyes (*29*) | Nominal | - Yes (n=134, 20.9%)<br>- No (n=266, 41.6%)<br>- N/A (n=239, 37.4%) | - Yes (n=120, 18.7%)<br>- No (n=325, 50.8%)<br>- N/A (n=194, 30.3%) |
| | Looking outside (*29*) | Nominal | - Yes (n=122, 19.0%)<br>- No (n=278, 43.5%)<br>- N/A (n=239, 37.4%) | - Yes (n=120, 18.7%)<br>- No (n=285, 44.6%)<br>- N/A (n=194, 30.3%) |
| | Looking inside (*29*) | Nominal | - Yes (n=58, 9.0%)<br>- No (n=342, 53.5%)<br>- N/A (n=239, 37.4%) | - Yes (n=80, 12.5%)<br>- No (n=365, 57.1%)<br>- N/A (n=194, 30.3%) |
| | Using mobile devices (*13*) | Nominal | - Yes (n=176, 27.5%)<br>- No (n=224, 35.0%)<br>- N/A (n=239, 37.4%) | - Yes (n=251, 39.2%)<br>- No (n=194, 30.3%)<br>- N/A (n=194, 30.3%) |
| | Watching in-vehicle display (*33*) | Nominal | - Yes (n=95, 14.8%)<br>- No (n=305, 47.7%)<br>- N/A (n=239, 37.4%) | - Yes (n=112, 17.5%)<br>- No (n=333, 52.1%)<br>- N/A (n=194, 30.3%) |
| | Handwriting (*13*) | Nominal | - Yes (n=20, 3.1%)<br>- No (n=380, 59.4%)<br>- N/A (n=239, 37.4%) | - Yes (n=24, 3.7%)<br>- No (n=421, 65.8%)<br>- N/A (n=194, 30.3%) |
| | Typing on laptop | Nominal | - Yes (n=27, 4.2%)<br>- No (n=373, 58.3%)<br>- N/A (n=239, 37.4%) | - Yes (n=29, 4.5%)<br>- No (n=416, 65.1%)<br>- N/A (n=194, 30.3%) |
| *Q16: In your most recent experience of MS (as a passenger) in an EV/FV, where were you seated?* | Seat (*13*) | Nominal | - Front passenger seat (n=82, 12.8%)<br>- Rear passenger seat (n=315, 49.2%)<br>- Not sure (n=3, 0.4%)<br>- N/A (n=239, 37.4%) | - Front passenger seat (n=89, 13.9%)<br>- Rear passenger seat (n=356, 55.7%)<br>- Not sure (n=0, 0%)<br>- N/A (n=194, 30.3%) |

*Note: N/A (not applicable) applies to those who have never traveled with EVs and FVs in the past one year, rendering the related question about their experience with EVs and FVs unnecessary. Additionally, it is also not applicable to those who have never encountered any MS in EVs or FVs within the past one year, thereby exempting them from responding to questions pertaining to their most recent MS experience in these vehicles.*

*Experience of MS in the past one year*
As illustrated in **Table 2**, respondents were asked about their experience of MS in EVs and FVs during the past year, including the frequency of riding EV or FV, and the frequency of experiencing MS in EV or FV. Further, we evaluated the severity level of MS in EVs and FVs by inquiring about details of their most recent MS experience in the past one year, including in-vehicle activities (Q15 & 16), road conditions (Q14), and driving behaviors in the drive (Q13). The severity of MS was assessed using a validated 4-point rating scale, which has been widely adopted in other MS-related studies (*27*) (*28*).

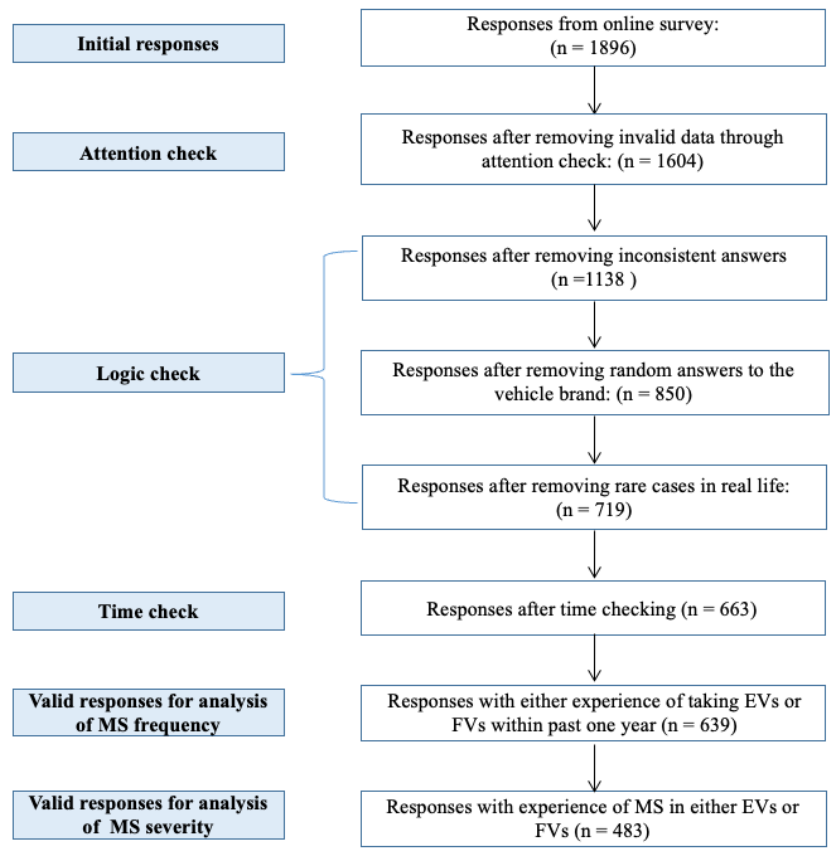

Figure 1: Data screening procedure

**Participants**
The questionnaire was distributed online in vehicle forums and on social media platforms (e.g., WeChat) in China. A total of 1896 respondents finished the questionnaire. To ensure data quality, attention checks, logic checks, and survey completion time checks were implemented (see **Figure 1**). Two simple questions served as attention checks when asking about the Experience of MS in EVs and FVs in part 3, leading to the exclusion of 292 respondents. In logic checks, 267 respondents who claimed to have never been passengers in EVs nor FVs but reported MS frequency (Q11) or severity (Q12) were removed; 66 respondents who claimed no experience of MS but reported severity (Q11) were excluded; 133 respondents who selected more than one choices for the seating position (Q16), such as having a seat in the front passenger seat and rear passenger seat at the same time were excluded; 288 respondents were removed as they provided mismatched vehicle brand and vehicle model information. Further exclusions were made for respondents who reported very rare life experiences, such as never taking a car or bus in the past ten years or before the age of 12 (131 respondents were excluded). Following the time check procedure proposed by Andereadis (*34*) and considering the similar reading speeds between Chinese characters and English words (*35*), a minimum threshold of 5 minutes was set for the study, which allowed for a response time of 1.4 seconds per question (43 questions in total) and a reading speed of 525 words per minute for the questionnaire. As a result of the time check, 54 respondents were excluded. Then, 24 respondents who did not take a ride in neither FV nor EV in the past one year were excluded, leading to 639 respondents being kept for final analysis, of which, 156 did not report MS in either FV or EV, 367 experienced MS in both FV and EV, 445 experienced MS in FV, and 400 experienced MS in EV in the past one year. Respondents who provided valid survey samples were compensated with 3 RMB. This study was approved by the Human and Artefacts Research Committee at the Hong Kong University of Science and Technology (HREP-2023-0111).

**Variable Extraction and Statistical analysis**
The extracted variables and their distribution (i.e., mean, standard deviation, minimum and maximum values) are summarized in **Table 1** and **Table 2**. Two ordinal logistic regression models considering the effect of repeated measures (i.e., respondent) were constructed using PROC GENMOD in SAS OnDemand for Academics, with the repeated measures being accounted for by Generalized Estimation Equation (*36*). In Model 1, we investigated the influential factors of MS frequency (Q11) based on their MS experience within the one past year. In the model, the odds ratio (OR) of experiencing MS more frequently was modeled. In Model 2, we investigated the influential factors of MS severity (Q12) based on the most recent MS experience within the last year. In the model, the odds ratio (OR) of experiencing more severe MS was modeled.

In all models, the vehicle type was included as the independent variable. Further, individual-related factors (Q1 – Q9) were used as independent variables in Model 1 and Model 2; and in-vehicle-activity-related (Q15 and Q16), road-condition-related (Q14), and driving-behaviors-related factors (Q13) were used as independent variables in Model 1. Before fitting the models, the correlations between all independent variables were assessed using the Spearman correlation (*37*). Independent variables with high absolute values of correlation coefficients (> 0.6) were either aggregated or abandoned (*38*). In our analysis, we initially constructed full models that included all independent variables, as well as their two-way interactions with vehicle type as predictors. Then, a backward stepwise selection process based on Quasi-likelihood under the Independence Criterion (QIC) (*39*) was adopted to select the independent variables. However, vehicle type was always kept in models, given that it is the variable of interest in this study. Post-hoc comparisons were conducted for all significant (*p*<.05) main effects and two-way interaction effects. All significant post-hoc contrasts were reported.

**RESULTS**
**Model 1: MS frequency**
**Table 3** presents the Wald statistics for the type 3 GEE analysis of model 1 (for the frequency of MS in the past one year). It was found that riding frequency was a significant predictor of MS frequency. At the same time, a significant interaction effect between vehicle type and FV ownership, and between vehicle type and MSSQ have been observed.

**Table 3: Wald statistics for type 3 GEE analysis of the final model 1 and model 2.**

| Model | Independent Variables | $\chi^2$-value | *P* |
|---|---|---|---|
| MS frequency (Model 1) | Vehicle type | $\chi^2(1) = 3.95.$ | .047* |
| | EV ownership | $\chi^2(1) = 2.13$ | .14 |
| | FV ownership | $\chi^2(1) = 31.50$ | < .001* |
| | MSSQ | $\chi^2(1) = 177.74$ | < .001* |
| | Riding frequency | $\chi^2(3) = 23.52$ | < .001* |
| | Vehicle type*FV ownership | $\chi^2(1) = 18.52$ | < .001* |

| | MSSQ*Vehicle type | $\chi^2(1) = 4.85$ | .03* |
|---|---|---|---|
| MS severity (Model 2) | Vehicle type | $\chi^2(1) = 7.11$ | .03* |
| | Gender | $\chi^2(1) = 4.83$ | .03* |
| | Crowded road | $\chi^2(1) = 6.37$ | .01* |
| | Uphill road | $\chi^2(1) = 16.89$ | < .001* |
| | Chatting | $\chi^2(1) = 5.46$ | .02* |
| | Listening to audio | $\chi^2(1) = 2.53$ | .1 |
| | Watching in-vehicle display | $\chi^2(1) = 24.03$ | < .001* |
| | Seat | $\chi^2(1) = 4.39$ | .11 |
| | EV ownership | $\chi^2(1) = 5.12$ | .02* |
| | FV ownership | $\chi^2(1) = 6.93$ | .009* |
| | Age | $\chi^2(1) = 7.59$ | .006* |
| | Sleep | $\chi^2(3) = 8.05$ | .045* |
| | MSSQ | $\chi^2(1) = 23.24$ | < .001* |
| | Riding frequency | $\chi^2(3) = 11.11$ | .01* |
| | Vehicle type*watching in-vehicle display | $\chi^2(1) = 3.27$ | .07* |
| | Vehicle type*EV ownership | $\chi^2(1) = 1.52$ | .2 |
| | Vehicle type*FV ownership | $\chi^2(1) = 3.72$ | .053 |
| | Age*vehicle type | $\chi^2(1) = 2.93$ | .09 |
| | MSSQ*vehicle type | $\chi^2(1) = 4.79$ | .03* |

Note: in this table, * marks significant results ($p<.05$).

Riding frequency was associated with the frequency of experiencing MS. especially those who took vehicles more frequently were more likely to experience a higher frequency of MS in vehicles, when drivers took a ride almost every day, they were more likely to experience MS more frequently, compared to that when they took a ride a few times a week (OR=2.00, 95%CI: [1.14, 3.53], $\chi^2(1)=5.89$, $p=.02$), a few times a month (OR=2.98, 95%CI: [1.68, 5.30], $\chi^2(1)=14.05$, $p=.0002$), and a few times a year or less (OR=3.50, 95%CI: [1.81, 6.72], $\chi^2(1)=14.08$, $p=.0002$). Those who took a ride a few times a week also reported having experienced MS more frequently compared to those who took a ride a few times a month (OR=1.49, 95%CI: [1.17, 1.90], $\chi^2(1)=10.31$, $p=.001$) and a few times a year or less (OR=1.74, 95%CI: [1.15, 2.65], $\chi^2(1)=6.73$, $p=.01$).

Regarding the interaction effect between vehicle type and FV ownership, it was found that for those who did not own an FV, taking a ride in FV was more likely to be associated with more frequent MS compared to that taking a ride in EV (OR=1.95, 95%CI: [1.38, 2.77], $\chi^2(1)=14.11$, $p=.0002$). Further, those who owned an FV were less likely to experience more frequent MS compared to those who did not own an FV, both when they took a ride in FV (OR=0.32, 95%CI: [0.24, 0.44], $\chi^2(1)=51.19$, $p<.0001$) and in EV (OR=0.67, 95%CI: [0.49, 0.92], $\chi^2(1)=5.95$, $p=.014$).

At the same time, regarding the interaction effect between MSSQ and vehicle type, it was found that MSSQ was positively associated with the frequency of MS in both EVs (OR=1.08, 95%CI: [1.07, 1.10], $\chi^2(1)=122.45$, $p<.0001$) and FVs (OR=1.10, 95%CI: [1.08, 1.11], $\chi^2(1)=163.44$, $p<.0001$). At the same time, for those with the same MSSQ score, taking an FV was more likely to lead to more frequent MS than taking an EV (OR=1.37, 95%CI: [1.03, 1.83], $\chi^2(1)=4.66$, $p=.03$).

**Model 2: MS severity**
**Table 3** also presents the Wald statistics for Model 2 (for the severity of the most recent experience of MS within the past one year). The following variables were found to be significantly associated with respondents' self-reported MS severity, i.e., *vehicle type, gender, crowded road, uphill road, chatting, listening to the audio, watching in-vehicle display, seat, EV ownership, FV ownership, age, sleep, MSSQ,* and *riding frequency*. Additionally, a significant interaction effect was observed between *vehicle type* and *MSSQ*.

First, road conditions and in-vehicle activities were associated with the severity of MS. Specifically, respondents who traveled through congested roads were less likely to experience more severe MS compared to those who did not (OR=0.66, 95%CI: [0.48, 0.91], $\chi^2(1)$ =6.37, $p$=.01). Conversely, traveling on uphill roads was positively correlated with more severity of MS (OR=3.73, 95%CI: [1.99, 6.99], $\chi^2(1)$ =16.89, $p$<.0001). Additionally, those who engaged in conversations with others were less likely to experience more severe MS than those who did not (OR=0.62, 95%CI: [0.41, 0.93], $\chi^2(1)$ =5.46, $p$=.02). Furthermore, those who watched in-vehicle display were less likely to experience more severe MS than those who did not (OR=0.37, 95%CI: [0.25, 0.55], $\chi^2(1)$ =24.03, $p$<.0001).

At the same time, individual differences also affected the severity of motion sickness. Compared to females, males were less likely to experience more severe MS (OR=0.68, 95%CI: [0.48, 0.96], $\chi^2(1)$ =4.83, $p$=.03). Age was negatively associated with the severity of MS: with every 1-year increase in age, the OR of experiencing MS in vehicles was 0.96, 95%CI: [0.94, 0.99], $\chi^2(1)$ =7.59, $p$=.006. Additionally, those who took a ride a few times per week were less likely to experience more severe MS than those who took a ride a few times a month (OR=0.56, 95%CI: [0.40, 0.79], $\chi^2(1)$ =10.95, $p$=.0009). Further, it seems that sleep time was negatively associated with the severity of MS. Specifcially, respondents who reported a routine sleep duration of fewer than 4 hours a day were less likely to experience severe MS in vehicles than those who slept 4 hours to 6 hours a day (OR=0.11, 95%CI: [0.01, 0.84], $\chi^2(1)$ =4.50, $p$=.03). Those who slept 6 hours a day were less likely to experience more severe MS than those slept 6 to 8 hours a day (OR=0.48, 95%CI: [0.24, 0.95], $\chi^2(1)$ =4.41, $p$=.03). Those who slept 6 to 8 hours a day were less likely to experience more severe MS than those who slept more than 8 hours a day (OR=0.52, 95%CI: [0.30, 0.93], $\chi^2(1)$ =4.89, $p$=.03).

Vehicle ownership was found to be associated with the MS as well. Those who reported having their EVs (OR=0.66, 95%CI: [0.46, 0.95], $\chi^2(1)$ =5.12, $p$=.02) and FVs (OR=0.63, 95%CI: [0.45, 0.89], $\chi^2(1)$ =6.93, $p$=.009) were associated with less likelihood of experiencing more severe MS than those who did not. Finally, the MSSQ, as expected, was positively associated with the MS severity, but to different extents in FV and EV (FV: OR=1.05, 95%CI: [1.03, 1.07], $\chi^2(1)$ =28.90, $p$<.0001; EV: OR=1.03, 95%CI: [1.01, 1.05], $\chi^2(1)$ =6.77, $p$=.01). At the same time, for the respondents with the same MSSQ score, those who took a FV was less likely to experience more severe MS than those who took an EV (OR=0.23, 95%CI: [0.08, 0.68], $\chi^2(1)$ =7.00, $p$=.008).

**DISCUSSION**
This study investigates whether EVs can cause more frequent and more severe MS than that of FVs. Using an online survey, 639 valid responses in China, the largest EV market, have been analyzed. First of all, in line with the findings in previous research (*20*), we found that the MSSQ was positively associated with the frequency and severity of experiencing MS. Additionally, as expected, we found that the MS frequency was positively associated with their frequency of taking a ride – those who took more rides were more likely to experience scenarios that may lead to MS.

Surprisingly, we found that the FV led to more frequent MS compared to EV, when controlling the MSSQ score of the respondents. This result is contradictory to the frequency of online complaints and common cognition of MS frequencies in FVs and EVs. The analysis of the MS severity may provide some insights into this phenomenon. We found that respondents also reported more severe MS in EVs compared to that in FVs when controlling the MSSQ score. Thus, it is possible that only those who experienced more severe MS preferred to share their unpleasant experience with others or online; while minor discomfort might be neglectable among the passengers. It is also possible that the MS inducers in FVs are common, but are the ones that are usually associated with minor MS.

Further, it seems that vehicle ownership also influenced both the MS frequency and MS severity. First, the FV ownership was found to be associated with the MS frequency. Those who owned FVs were less likely to experience MS in either FVs or EVs compared to those who did not. At the same time, those who did not own FVs seemed to experience more frequent MS in FVs compared to that in EVs. While the

*Weiyin Xie, Chunxi Huang, Jiyao Wang, and Dengbo He*

MS severity was associated with both EV ownership and FV ownership with those who owned EVs or FVs reporting less likelihood of experiencing more severe MS. This result can be interpreted in several ways. First, those who owned a vehicle may have higher frequency of taking a ride of the vehicle. Given that the training program is effective in reducing the pilots' resistance to MS in fighter jets (*40*), more frequent vehicle riding/riding experience may also have trained the vehicle owners so that they became less sensitive to the MS inducers in vehicles. It is also possible that those who owned a vehicle were also drivers and they may have better capability to anticipate the motion of the vehicles even when they were passengers in the vehicle (*16*). However, only FV ownership was associated with less frequent MS frequency. It is possible that certain inducers of MS inducers are there in the FV but not in EV, so the passengers/owners were only desensitized to these inducers. For example, it is possible that the unpleasant odors that cause minor MS (*13*) only exist in FVs.

Our study also found that the MS severity was related to road conditions and passengers' in-vehicle activities. The road conditions that we identified to be associated with MS severity include traffic density (i.e., congested or not) and road slope. These findings suggest that the magnitude and directions of the vehicle acceleration are inducers of motion sickness. For example, the misalignment of the head and the gravito-inertial force while driving uphill may cause MS (*32*). Further, although the congested road is associated with a more frequent stop-and-go driving scenario, the congested roads may have constrained the velocity and acceleration of the vehicle, leading to reduced MS symptoms. Also, certain in-vehicle activities were found to be associated with MS severity. This highlights the importance of further quantifying the weights of different MS inducers in vehicles. Our results also suggest that chatting with others can alleviate MS symptoms. This is not surprising, as chatting can shift passengers' attention away from the MS-related symptoms and alleviate the perceived MS severity level (*41*). However, contrary to previous research suggesting that looking at displays is more likely to lead to MS compared to focusing on the road outside (31)(33), our results suggest that watching the in-vehicle display was associated with less severe MS symptoms. It is possible that only those who were less sensitive to MS would engage in in-vehicle activities such as watching an in-vehicle screen. Future empirical studies are needed to better explain this finding.

We also found that MS severity is associated with individual differences. In line with previous studies (*21*), we found that older passengers tended to experience less severe MS. Similarly, it is not surprising to find that female passengers experienced more severe MS compared to male passengers, given that female was found to be more susceptible to MS than male (*22*). Lastly, we found that daily living habit was also a predictor of MS severity among passengers, with those who slept more being more likely to experience more severe MS. This is contradictory to previous studies which found that inadequate sleep can lead to more MS (27). However, it should be noted that the "adequate sleep" might be different for different people. In other words, short sleep is not equal to inadequate sleep. Further, self-reported sleep may not be equivalent to true sleep duration being monitored in a lab setting. Future research is still needed to better reveal the relationship among the required sleep time, deprivation of sleep, and sensitivity to MS in vehicles.

Finally, it should be noted that, although we tried to control the quality of the responses obtained from the study using attentional check and logical checks, and only focused on respondents' riding experience in the past one year, or in their most recent MS experience, the readers should still be cautious about the validity of our findings. For example, it is possible that those who have complaints about the MS in FV or EV were more preferred to participate in MS-related studies and thus introduce bias into the results of our survey-based study. Thus, future empirical experiments are needed to further validate the findings obtained in our study. Further, the MS is highly related to the driving style of the drivers and the in-vehicle environment (e.g., temperature (*13*)). Although we tried to evaluate the driving style in the questionnaire, other unconsidered characteristics of the vehicle dynamics might affect MS as well. Though the large sample size in our study may partially cancel out the influence of the unconsidered covariants, more strictly designed on-road or observational studies are still needed. Overall, though with the above-mentioned limitations, this survey study has provided valuable insights into the direction of future empirical studies on in-vehilce MS.

*Weiyin Xie, Chunxi Huang, Jiyao Wang, and Dengbo He*

**CONCLUSIONS**
Based on 639 valid survey responses from EVs and FVs passengers, for the first time, we systematically investigated the occurrence of MS in EVs and FVs. Our results suggest that in general, when controlling the respondents' susceptibility to motion-induced MS, passengers were more likely to experience MS in FVs compared to that in EVs; but less severe MS symptoms were observed in FVs compared to that in EVs. We further found that MS severity was associated with road conditions, in-vehicle activities, and individual differences. These findings highlight the need to consider the unique characteristics of EVs when investigating the causes of MS, as the inducers of MS in EVs may differ from those in FVs. Given that most existing research on MS in vehicles focused on the characteristics of FVs or have been conducted in FVs, our study indicates it is necessary to differentiate and investigate the potential MS inducers in EVs and explore countermeasures to reduce MS in EVs. Future studies are should further quantify the inducers of the MS in EVs to better explain the findings in our study.

**ACKNOWLEDGMENTS**
This research is supported by the Guangzhou-HKUST(GZ) Joint Funding Scheme (No. 2023A03J00073), and partially by the Guangzhou Municipal Science and Technology Project (No. 2023A03J0011), and the Guangzhou Science and Technology Program City-University Joint Funding Project (No. 2023A03J0001).

**AUTHOR CONTRIBUTIONS**
The authors confirm their contribution to the paper as follows: study conception and design: Weiyin Xie, Dengbo He; data collection: Weiyin Xie, Jiyao Wang; analysis and interpretation of results: Weiyin Xie, Chunxi Huang, Dengbo He; draft manuscript preparation: Weiyin Xie, Chunxi Huang, Jiyao Wang, Dengbo He. All authors reviewed the results and approved the final version of the manuscript.

**REFERENCES**
1. China Energy-Saving Vehicle & NEV Roadmap 2.0: Curbing Carbon Emissions for a Green Society - MarkLines Automotive Industry Portal. https://www.marklines.com/en/report/rep2142_202104. Accessed Jun. 23, 2023.

2. Team, T. W. How Do Electric Vehicles Compare To Gas Cars? | Wallbox. New, Jan 04, 2021.

3. Smyth, J., J. Robinson, R. Burridge, P. Jennings, and R. Woodman. Towards the Management and Mitigation of Motion Sickness–An Update to the Field. 2021.

4. Turner, M., M. J. Griffin, and I. Holland. Airsickness and Aircraft Motion during Short-Haul Flights. *Aviation, space, and environmental medicine*, Vol. 71, No. 12, 2000, pp. 1181–1189.

5. SUZUKI, H., H. SHIROTO, and K. TEZUKA. Effects of Low Frequency Vibration on Train Motion Sickness. *Quarterly Report of RTRI*, Vol. 46, No. 1, 2005, pp. 35–39.

6. Chang, E., H. T. Kim, and B. Yoo. Virtual Reality Sickness: A Review of Causes and Measurements. *International Journal of Human–Computer Interaction*, Vol. 36, No. 17, 2020, pp. 1658–1682.

7. Saruchi, S. A., M. H. M. Ariff, H. Zamzuri, M. A. A. Rahman, N. Wahid, N. Hassan, N. A. Izni, F. Yakub, N. A. Husain, and K. A. Kassim. Motion Sickness Mitigation in Autonomous Vehicle: A Mini-Review. *Journal of the Society of Automotive Engineers Malaysia*, Vol. 5, No. 2, 2021, pp. 260–272.

8. Li, D., and L. Chen. *Mitigating Motion Sickness in Automated Vehicles with Vibration Cue System*. 2022.

9. Sato, H., Y. Sato, A. Takamatsu, M. Makita, and T. Wada. Earth-Fixed Books Reduce Motion Sickness When Reading With a Head-Mounted Display. *Frontiers in Virtual Reality*, Vol. 3, 2022, p. 909005. https://doi.org/10.3389/frvir.2022.909005.

10. Do Electric Cars Cause Motion Sickness? (More or Less) - Rechargd. https://rechargd.com/do-electric-cars-cause-motion-sickness/. Accessed Jun. 23, 2023.

11. Ozaki, R., and K. Sevastyanova. Going Hybrid: An Analysis of Consumer Purchase Motivations. *Energy policy*, Vol. 39, No. 5, 2011, pp. 2217–2227.

12. Reason, J. T., and J. J. Brand. *Motion Sickness / J. T. Reason, J. J. Brand.* Academic Press, London ; New York, 1975.

13. Schmidt, E. A., O. X. Kuiper, S. Wolter, C. Diels, and J. E. Bos. An International Survey on the Incidence and Modulating Factors of Carsickness. *Transportation research part F: traffic psychology and behaviour*, Vol. 71, 2020, pp. 76–87.

14. Zamzuri, H., N. Hassan, and M. H. M. Ariff. Modeling of Head Movements towards Lateral Acceleration Direction via System Identification for Motion Sickness Study. 2018.

15. Golding, J. F., and M. A. Gresty. Motion Sickness. *Current Opinion in Neurology*, Vol. 18, No. 1, 2005, pp. 29–34. https://doi.org/10.1097/00019052-200502000-00007.

16. Rolnick, A., and R. E. Lubow. Why Is the Driver Rarely Motion Sick? The Role of Controllability in Motion Sickness. *Ergonomics*, Vol. 34, No. 7, 1991, pp. 867–879.

17. Car Sickness. https://tesla-info.com/tips/car-sickness.php. Accessed Jun. 23, 2023.

18. Garcia, C. B., and F. J. Gould. Survivorship Bias. *Journal of Portfolio Management*, Vol. 19, No. 3, 1993, p. 52.

19. Demand for Electric Cars Is Booming, with Sales Expected to Leap 35% This Year after a Record-Breaking 2022 - News. *IEA*. https://www.iea.org/news/demand-for-electric-cars-is-booming-with-sales-expected-to-leap-35-this-year-after-a-record-breaking-2022. Accessed Jul. 20, 2023.

20. Golding, J. F. Motion Sickness Susceptibility Questionnaire Revised and Its Relationship to Other Forms of Sickness. *Brain Research Bulletin*, Vol. 47, No. 5, 1998, pp. 507–516. https://doi.org/10.1016/s0361-9230(98)00091-4.

21. Jones, M. L., V. C. Le, S. M. Ebert, K. H. Sienko, M. P. Reed, and J. R. Sayer. Motion Sickness in Passenger Vehicles during Test Track Operations. *Ergonomics*, Vol. 62, No. 10, 2019, pp. 1357–1371.

22. Jokerst, M. D., M. Gatto, R. Fazio, P. J. Gianaros, R. M. Stern, and K. L. Koch. Effects of Gender of Subjects and Experimenter on Susceptibility to Motion Sickness. *Aviation, space, and environmental medicine*, Vol. 70, No. 10, 1999, pp. 962–965.

23. Caillet, G., G. Bosser, G. C. Gauchard, N. Chau, L. Benamghar, and P. P. Perrin. Effect of Sporting Activity Practice on Susceptibility to Motion Sickness. *Brain research bulletin*, Vol. 69, No. 3, 2006, pp. 288–293.

24. Kaplan, J., J. Ventura, A. Bakshi, A. Pierobon, J. R. Lackner, and P. DiZio. The Influence of Sleep Deprivation and Oscillating Motion on Sleepiness, Motion Sickness, and Cognitive and Motor Performance. *Autonomic neuroscience*, Vol. 202, 2017, pp. 86–96.

25. Lawther, A., and M. J. Griffin. A Survey of the Occurrence of Motion Sickness amongst Passengers at Sea. *Aviation, space, and environmental medicine*, Vol. 59, No. 5, 1988, pp. 399–406.

26. Golding, J. F., O. Prosyanikova, M. Flynn, and M. A. Gresty. The Effect of Smoking Nicotine Tobacco versus Smoking Deprivation on Motion Sickness. *Autonomic Neuroscience*, Vol. 160, No. 1–2, 2011, pp. 53–58.

27. Turner, M. Motion Sickness in Public Road Transport: Passenger Behaviour and Susceptibility. *Ergonomics*, Vol. 42, No. 3, 1999, pp. 444–461.

28. LAWTHER, A., and M. J. GRIFFIN. The Motion of a Ship at Sea and the Consequent Motion Sickness amongst Passengers. *Ergonomics*, Vol. 29, No. 4, 1986, pp. 535–552. https://doi.org/10.1080/00140138608968289.

29. Kato, K., and S. Kitazaki. A Study of Carsickness of Rear-Seat Passengers Due to Acceleration and Deceleration When Watching an in-Vehicle Display. *Review of Automotive Engineering*, Vol. 27, 2006, pp. 465–469.

30. Golding, J. F., A. G. Mueller, and M. A. Gresty. A Motion Sickness Maximum around the 0.2 Hz Frequency Range of Horizontal Translational Oscillation. *Aviation, space, and environmental medicine*, Vol. 72, No. 3, 2001, pp. 188–192.

31. DiZio, P., J. Ekchian, J. Kaplan, J. Ventura, W. Graves, M. Giovanardi, Z. Anderson, and J. R. Lackner. An Active Suspension System for Mitigating Motion Sickness and Enabling Reading in a Car. *Aerospace medicine and human performance*, Vol. 89, No. 9, 2018, pp. 822–829.

32. Golding, J. F., W. Bles, J. E. Bos, T. Haynes, and M. A. Gresty. Motion Sickness and Tilts of the Inertial Force Environment: Active Suspension Systems vs. Active Passengers. *Aviation, space, and environmental medicine*, Vol. 74, No. 3, 2003, pp. 220–227.

33. Brietzke, A., R. Pham Xuan, A. Dettmann, and A. C. Bullinger. Influence of Dynamic Stimulation, Visual Perception and Individual Susceptibility to Car Sickness during Controlled Stop-and-Go Driving. *Forschung im Ingenieurwesen*, Vol. 85, No. 2, 2021, pp. 517–526. https://doi.org/10.1007/s10010-021-00441-6.

34. Andreadis, I. Data Quality and Data Cleaning. *Matching voters with parties and candidates. Voting advice applications in comparative perspective*, 2014, pp. 79–91.

35. Trauzettel-Klosinski, S., K. Dietz, and Ir. S. Group. Standardized Assessment of Reading Performance: The New International Reading Speed Texts IReST. *Investigative ophthalmology & visual science*, Vol. 53, No. 9, 2012, pp. 5452–5461.

36. SAS Help Center: The GENMOD Procedure. .

37. Wissler, C. The Spearman Correlation Formula. *Science*, Vol. 22, No. 558, 1905, pp. 309–311.

38. Dormann, C. F., J. Elith, S. Bacher, C. Buchmann, G. Carl, G. Carré, J. R. G. Marquéz, B. Gruber, B. Lafourcade, and P. J. Leitão. Collinearity: A Review of Methods to Deal with It and a Simulation Study Evaluating Their Performance. *Ecography*, Vol. 36, No. 1, 2013, pp. 27–46.

39. 26100 - QIC Goodness of Fit Statistic for GEE Models. https://support.sas.com/kb/26/100.html. Accessed Oct. 10, 2023.

40.     Eslami, R., H. Mohsenzadeh, and M. Yari. Investigating the Effect of a Selected Exercise Training Course on the Motion Sickness in Pilots and Flight Crew Members of the Army of the Islamic Republic of Iran. *Journal of Military Medicine*, Vol. 24, No. 9, 2023, pp. 1566–1578.

41.     Bos, J. E. Less Sickness with More Motion and/or Mental Distraction. *Journal of Vestibular Research*, Vol. 25, No. 1, 2015, pp. 23–33.